\documentclass[aps,prl,twocolumn,superscriptaddress]{revtex4-1}
\usepackage{graphicx}
\usepackage{epsfig}
\usepackage{amsmath,amsfonts,amsthm}
\usepackage{amssymb}
\usepackage{times}
\usepackage{braket}
\usepackage{units}
\usepackage{multirow}
\usepackage{mathrsfs}
\usepackage{theorem}
\usepackage{bm}
\usepackage{threeparttable}
\usepackage{url}
\usepackage[T1]{fontenc}
\usepackage{csquotes}
\usepackage{natbib}
\usepackage{float}
\usepackage{appendix}
\usepackage[colorlinks=true,linkcolor=blue,citecolor=blue,urlcolor=blue]{hyperref}
\usepackage{bookmark}
\usepackage{soul}
\MakeOuterQuote{"}
\usepackage{dcolumn}

\begin{document}

\title{Experimental creation of Multi-Photon High-Dimensional Layered Quantum States}
\author{Xiao-Min Hu}
\affiliation{CAS Key Laboratory of Quantum Information, University of Science and Technology of China, Hefei, 230026, People's Republic of China}
\affiliation{CAS Center For Excellence in Quantum Information and Quantum Physics, University of Science and Technology of China, Hefei, 230026, People's Republic of China}

\author{Wen-Bo Xing}
\affiliation{CAS Key Laboratory of Quantum Information, University of Science and Technology of China, Hefei, 230026, People's Republic of China}
\affiliation{CAS Center For Excellence in Quantum Information and Quantum Physics, University of Science and Technology of China, Hefei, 230026, People's Republic of China}

\author{Chao Zhang}
\affiliation{CAS Key Laboratory of Quantum Information, University of Science and Technology of China, Hefei, 230026, People's Republic of China}
\affiliation{CAS Center For Excellence in Quantum Information and Quantum Physics, University of Science and Technology of China, Hefei, 230026, People's Republic of China}

\author{Bi-Heng Liu}
\email{bhliu@ustc.edu.cn}
\affiliation{CAS Key Laboratory of Quantum Information, University of Science and Technology of China, Hefei, 230026, People's Republic of China}
\affiliation{CAS Center For Excellence in Quantum Information and Quantum Physics, University of Science and Technology of China, Hefei, 230026, People's Republic of China}

\author{Matej Pivoluska}
\email{pivoluskamatej@gmail.com}
\affiliation{Institute of Computer Science, Masaryk University, Botanick\'{a} 68a, 60200 Brno, Czech Republic}
\affiliation{Institute of Physics, Slovak Academy of Sciences,
D\'{u}bravsk\'{a} cesta 9,
845 11 Bratislava, Slovakia}

\author{Marcus Huber}
\email{marcus.huber@univie.ac.at}
\affiliation{Institute for Quantum Optics and Quantum Information (IQOQI), Austrian Academy of Sciences, A-1090 Vienna, Austria}

\author{Yun-Feng Huang}
\affiliation{CAS Key Laboratory of Quantum Information, University of Science and Technology of China, Hefei, 230026, People's Republic of China}
\affiliation{CAS Center For Excellence in Quantum Information and Quantum Physics, University of Science and Technology of China, Hefei, 230026, People's Republic of China}

\author{Chuan-Feng Li}
\email{cfli@ustc.edu.cn}
\affiliation{CAS Key Laboratory of Quantum Information, University of Science and Technology of China, Hefei, 230026, People's Republic of China}
\affiliation{CAS Center For Excellence in Quantum Information and Quantum Physics, University of Science and Technology of China, Hefei, 230026, People's Republic of China}

\author{Guang-Can Guo}
\affiliation{CAS Key Laboratory of Quantum Information, University of Science and Technology of China, Hefei, 230026, People's Republic of China}
\affiliation{CAS Center For Excellence in Quantum Information and Quantum Physics, University of Science and Technology of China, Hefei, 230026, People's Republic of China}

\date{\today}

\begin{abstract}
Quantum entanglement is one of the most important resources in quantum information. In recent years, the research of quantum entanglement mainly focused on the increase in the number of entangled qubits or the high-dimensional entanglement of two particles. Compared with qubit states, multipartite high-dimensional entangled states have beneficial properties and are powerful for constructing quantum networks. However, there are few studies on multipartite high-dimensional quantum entanglement due to the difficulty of creating such states. In this paper, we experimentally prepared a multipartite high-dimensional state $|\Psi_{442}\rangle=\frac{1}{2}(|000\rangle+|110\rangle+|221\rangle+|331\rangle)$ by using the path mode of photons. We obtain the fidelity $F=0.854\pm0.007$ of the quantum state, which proves a real multipartite high-dimensional entangled state. Finally, we use this quantum state to demonstrate a layered quantum network in principle. Our work highlights another route towards complex quantum networks.
\end{abstract}

\maketitle

Quantum entanglement \cite{Ryszard,Review}, as one of the most important phenomena in quantum information, has been proven to play a central role in many applications: fault-tolerant quantum computation \cite{Shor}, device-independent quantum communication \cite{Hillery} and quantum precision measurements \cite{Vittorio}. In recent years, the research on quantum entanglement mainly focused on multipartite qubit systems \cite{Zhong,Wang}, or two-partite high-dimensional systems \cite{Bavaresco,Martin}. For example, in optical systems, the preparation of quantum entanglement mainly develops in two directions: one is to increase the number of qubits of entanglement, such as 12-photon entanglement \cite{Zhong}, 18 qubit entanglement \cite{Wang}, the other is to increase the dimensionality of two photons, such as entanglement of $100\times100$ orbital angular momentum (OAM) degrees of freedom \cite{Krenn2014}. Since the higher-dimensional entanglement is naturally present in down-conversion processes, it would be desirable to harness this high-dimensionality for multi-photon experiments. Unfortunately, there are few experimental studies on multipartite high-dimensional entangled quantum states. The main reason is that the preparation of such entangled states requires very delicate manipulation for high-dimensional quantum systems.

\begin{figure}[tbph]
\begin{center}
\includegraphics [width= 1\columnwidth]{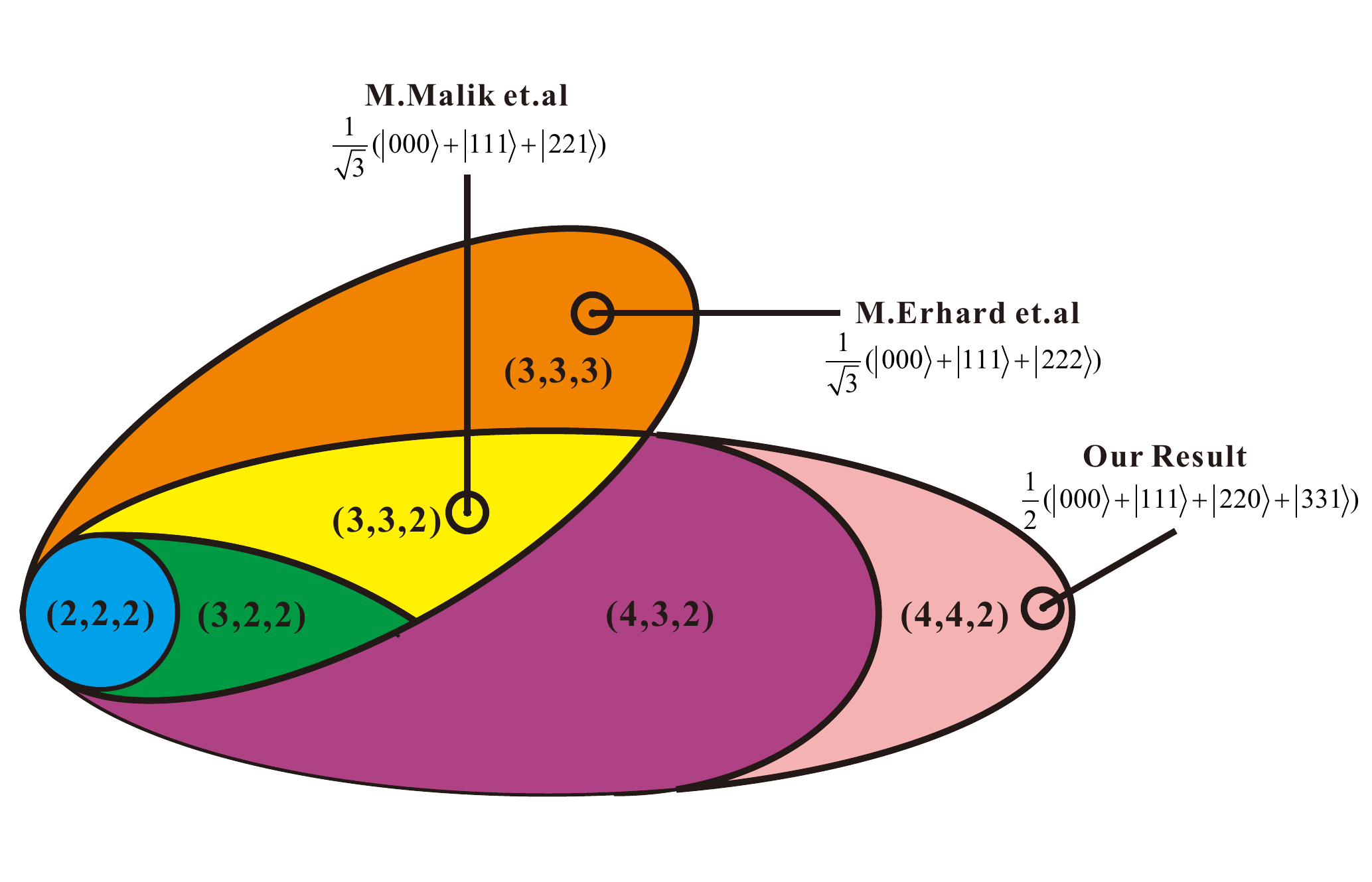}
\end{center}
\caption{Schematic representation of a few sets of states with a given Schmidt number vector \cite{Huber}. Multipartite states can be classified by calculating Schmidt number vectors of each particle in a multipartite state. After years of development, the preparation of (2, 2, 2), (3, 3, 2) \cite{Malik} and (3, 3, 3) \cite{Erhard} quantum states has been completed. Here we have completed the preparation of (4, 4, 2) quantum state for the first time. }
\label{fig1}
\end{figure}

In the field of quantum information, the most commonly used high-dimensional degrees of freedom (DoFs) in photonic systems are: orbital angular momentum (OAM) \cite{Mair}, time bin \cite{Marcikic}, path \cite{Boschi,Kim03,Hu2016,Jianwei}. The high-dimensional DoFs of these photons have their advantages and disadvantages in the application of quantum information. For example, using OAM is easier to expand the dimension \cite{Robert}, but the fidelity of the preparation and operation is lower \cite{Feiran} and the long-distance distribution is more difficult \cite{Sit}. The advantage of time bin DoF is that it is more suitable for long-distance distribution \cite{Takuya}, however, it is difficult to implement arbitrary unitary operations on time bins. The path DoF has a very high fidelity and is easy to manipulate \cite{Hu1}, but its dimension scalability is still a challenge and long-distance distribution is also difficult \cite{Hu2}. Until now, multipartite high-dimensional quantum entangled states have been successfully prepared only on OAM DoF. If classified according to Schmidt number vector \cite{Huber}, (3, 3, 2) state ($1/\sqrt{3}(|000\rangle+|111\rangle+|221\rangle)$) \cite{Malik}, (3, 3, 3) state ($1/\sqrt{3}(|000\rangle+|111\rangle+|222\rangle)$) \cite{Erhard} and high dimensional Dicke states \cite{Hiesmayr} have been successfully prepared on OAM (See Fig. \ref{fig1}). Due to the difficulties of state preparation, the largest dimension encoded in each photon is 3, and the observed fidelities are a bit low compared to other multi-photon experiments.

For potential application in quantum key distribution, Ref. \cite{Huber,Pivoluska} proposed to use a multipartite high-dimensional quantum state, that has so far not been created in any experiment:
\begin{equation}
|\Psi_{442}\rangle=\frac{1}{2}(|000\rangle+|111\rangle+|220\rangle+|331\rangle).
\end{equation}

\begin{figure*}[tbph]
\begin{center}
\includegraphics [width= 1.8\columnwidth]{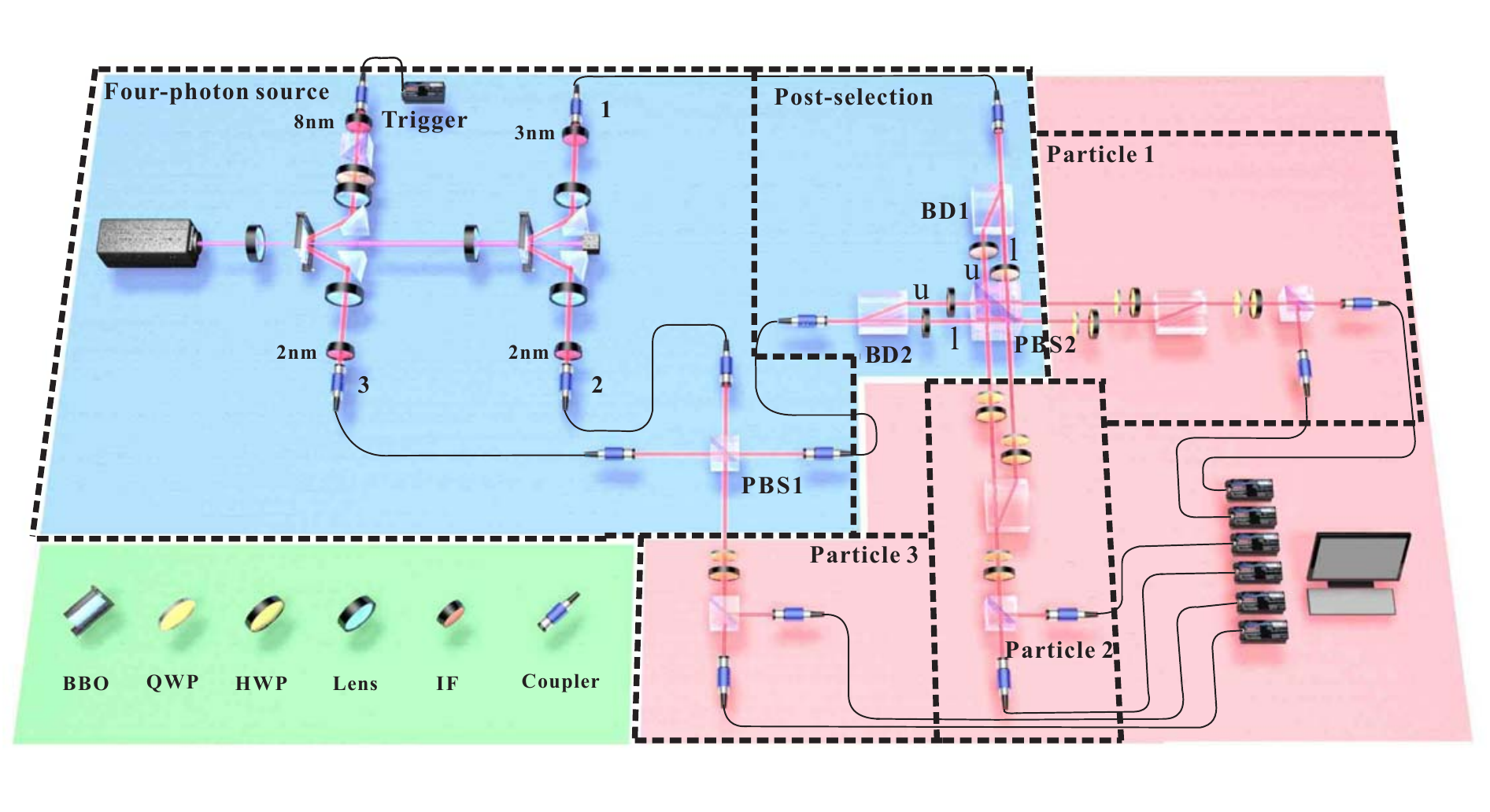}
\end{center}
\caption{Experimental setup for three-partite layered quantum entangled states. The blue region shows the process of preparing quantum states. The ultraviolet pulse laser (390~nm) from the doubler sequentially pumps nonlinear crystals to generate two entanglement source $(|HH\rangle+|VV\rangle)/\sqrt{2}$ at 780~nm. One of the photons serves as a trigger, the other three photons are prepared on a three-photon GHZ state $(|HHH\rangle+|VVV\rangle)/\sqrt{2}$ \cite{Zhang}. Finally, the particle 3 is directly measured in the polarization basis, the other two photons are incident into the high-dimensional post-selection device, which is used to post-select the two-dimensional polarization entangled state $(|HH\rangle+|VV\rangle)/\sqrt{2}$ into the polarization-path hybrid entangled state $(|H_{u}H_{u}\rangle+|V_{u}V_{u}\rangle+|H_{l}H_{l}\rangle+|V_{l}V_{l}\rangle)/2$. If we encode $|H_{u}\rangle\rightarrow|0\rangle$, $|V_{u}\rangle\rightarrow|2\rangle$, $|H_{l}\rangle\rightarrow|1\rangle$, and $|V_{l}\rangle\rightarrow|3\rangle$, we will get three-partite layered entangled state $(|000\rangle+|111\rangle+|220\rangle+|331\rangle)/2$. The pink part shows the measurement device for three photon states. Among them, particle 1 and 2 are encoded on 4-dimensional space, and particle 3 is encoded on 2-dimensional space.}
\label{fig2}
\end{figure*}


Notice that the first two photons, A and B, live in a four-dimensional space, whereas the third photon, C, lives in a two-dimensional space. The state's dimensionality is given by a vector of three numbers (4, 4, 2), which are the ordered ranks of the single particle reductions of the state density operator:
\begin{equation}
\operatorname{rank}\left(\rho_{A}\right)=4, \quad\operatorname{rank}\left(\rho_{B}\right)=4, \quad\operatorname{rank}\left(\rho_{C}\right)=2,
\end{equation}
where $\rho_{i}=\operatorname{Tr}_{\overline{i}} | \Psi \rangle_{442}\left\langle\left.\Psi\right|_{442}\right.$. This quantum state is obviously different from the general GHZ state ($|\Psi_{333}\rangle$), and it contains quantum state $|\Psi_{332}\rangle$, see Fig. \ref{fig1}.

This quantum state exhibits different properties from other multipartite entangled states. If we observe the two-dimensional subspaces of this quantum state, we find that there is a perfect correlation between particle A, B and C in $\{|0\rangle, |1\rangle\}$ space ($\{|000\rangle, |111\rangle\}$), simultaneously there is also a perfect correlation between A, B and C in $\{|2\rangle, |3\rangle\}$ space and C in $\{|0\rangle, |1\rangle\}$ space ($\{|220\rangle, |331\rangle\}$). So there is always a perfect GHZ correlation between A, B, and C if we observe the quantum state in a specific subspace. On the other hand, if C is detected in mode $|0\rangle$, then A, B are perfectly correlated in modes $|0\rangle$ and $|2\rangle$; if C is detected in mode $|1\rangle$, then A, B are perfectly correlated in modes $|1\rangle$ and $|3\rangle$. This property is quite different from all the previous states \cite{Malik,Erhard,Hiesmayr}, and enables a layered quantum network and exhibits the advantage of high-dimensional systems. For this reason this quantum state has been called layered quantum state. In this letter, we use the path and polarization DoFs of photons to build four-dimensional systems, demonstrate the creation and verification of one such entangled state with a fidelity of 0.854, which is higher than previous experiments \cite{Malik,Erhard,Hiesmayr}. Owing to the highest fidelity, we demonstrate its application in a highly efficient layered quantum communication protocol in principle. This technique can be applied to construct complex quantum networks in the future.

\begin{figure*}[tbph]
\begin{center}
\includegraphics [width= 1.8\columnwidth]{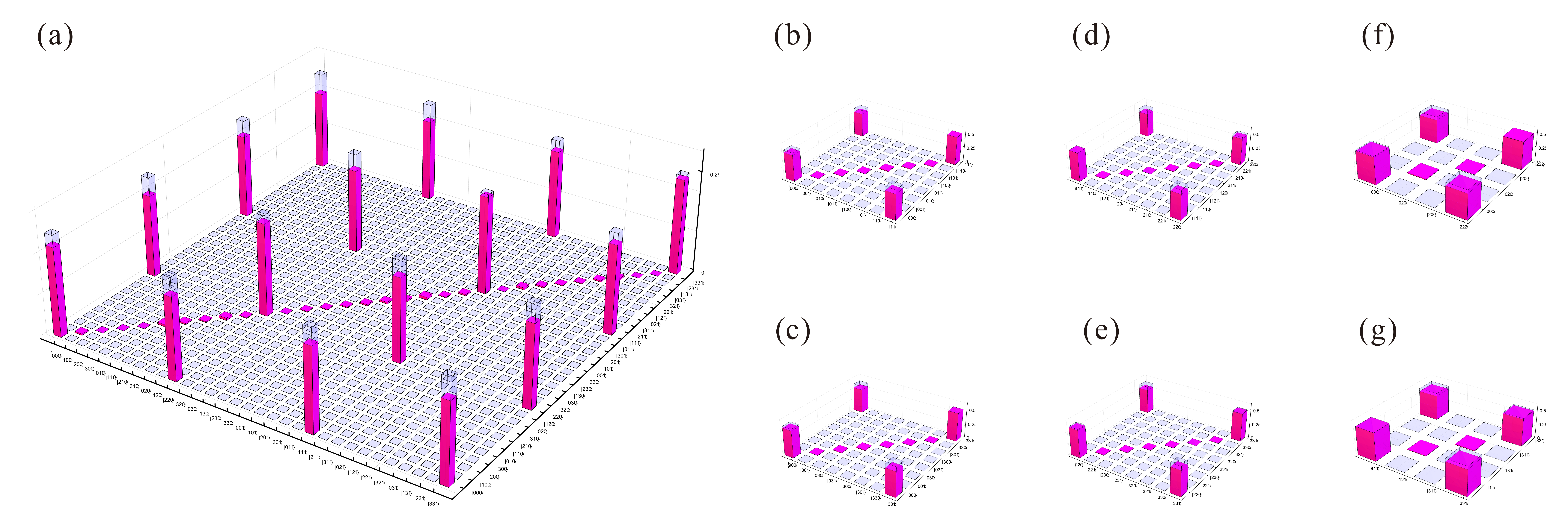}
\end{center}
\caption{(a) The 32 diagonal elements and 6 unique real parts of off-diagonal elements of $\rho_{exp}$ (elements which were not measured in the experiment are left grey). From these values, we can calculate $F_{exp}= 0.854\pm0.007$. The fidelity exceeds the upper bound $0.75$ with 14 standard deviations, which proves that we have successfully prepared (4, 4, 2) entangled state. The red column represents the experimental value and the transparent column represents the theoretical value of an ideal state. The difference is due to imperfect interferences and multi-photon noise. (b)-(g) Diagonal and real parts of unique off-diagonal elements of GHZ state in two-dimensional subspaces. From these values, we can get the fidelities of these states.
(b)$F((|000\rangle+|111\rangle)/\sqrt{2})=0.910\pm0.029$, (c)$F((|000\rangle+|331\rangle)/\sqrt{2})=0.906\pm0.030$, (d) $F((|111\rangle+|220\rangle)/\sqrt{2})=0.914\pm0.030$, (e) $F((|220\rangle+|331\rangle)/\sqrt{2})=0.922\pm0.028$, (f) $F((|000\rangle+|220\rangle)/\sqrt{2})=0.933\pm0.030$, (g) $F((|111\rangle+|331\rangle)/\sqrt{2})=0.941\pm0.031$. The fidelities of these states are far beyond the lower bound of genuine multipartite entanglement ($F=0.5$).}
\label{fig3}
\end{figure*}

Because the polarization DoF of photons has only two levels, it is impossible to construct high-dimensional quantum states only by using polarization DoF in a single photon \cite{Bogdanov,Kim}. So we use a beam displacer (BD) to additionally use the path DoF and combine it with the polarization DoF to complete hybrid high-dimensional coding.

First, we prepare a standard three-photon qubit GHZ state, as shown in Fig. \ref{fig2}. The ultraviolet pulse laser (390~nm) from the doubler sequentially pumps two entanglement source (780~nm). Then, the two entanglement sources, both are prepared in the state $(|HH\rangle+|VV\rangle)/\sqrt{2}$. Afterwards, the output photon 2 and 3 are directed to a polarizing beam splitter (PBS1). Here, all the PBSs are set to transmit horizontal polarization (H) and reflect vertical polarization (V). If there is one and only one photon in each output of the four-photon source part, the state $(|HHH\rangle+|VVV\rangle)/\sqrt{2}$ is generated (one of the photons acts as a trigger) \cite{Zhang}.

Then, one photon is directly measured in the polarization basis (particle 3), and the other two photons incident to a post-selection setup consisting of BD1, 2, PBS2 and half-wave plates (HWPs) set at $22.5^{\circ}$. At this time, we define the polarization DoF of particle 1 and particle 2 in the upper path as $|H\rangle\longrightarrow|0\rangle$, $|V\rangle\longrightarrow|2\rangle$, and lower path $|H\rangle\longrightarrow|1\rangle$ and $|V\rangle\longrightarrow|3\rangle$. The function of this device is to post-select the two-dimensional entangled state $(|HH\rangle+|VV\rangle)/\sqrt{2}$ into the four-dimensional entangled state $(|00\rangle+|11\rangle+|22\rangle+|33\rangle)/2$ \cite{Boschi,Ciampini,Hu3,Supple}.

After this post-selection, the quantum state becomes:
\begin{equation}
\begin{split}
|\Psi_{442}\rangle&=\frac{1}{2}(|00\rangle+|22\rangle)|0\rangle+\frac{1}{2}(|11\rangle+|33\rangle)|1\rangle \\
&=\frac{1}{2}(|000\rangle+|111\rangle+|220\rangle+|331\rangle).
\end{split}
\end{equation}


\begin{table*}[htbp]
\caption{Security analysis for QKD in different subspaces}
\begin{tabular}{|c|c|c|c|c|c|}
  \hline
  $Subspace$ & $QBER_{Z}$ &  $QBER_{X}$ & $QBER_Z(AB)$& $QBER_Z(AC)$&\textit{Key per round}   \\
  \hline
    $|000\rangle,|111\rangle$ & $0.044\pm0.009$ & $0.069\pm0.005$ & $0.023 \pm 0.006$
& $0.033 \pm 0.008$  & $0.428$  \\
  \hline
  $|220\rangle,|331\rangle$ & $0.023\pm0.006$ & $0.066\pm0.005$ & $0.014 \pm 0.005$
& $0.013 \pm 0.005$ &  $0.524$  \\
  \hline
  $|00\rangle,|22\rangle$ & $0.015\pm0.005$ & $0.061\pm0.010$ & $ $ & $ $& $0.508$
   \\
  \hline
  $|11\rangle,|33\rangle$ & $0.012\pm0.005$ &$0.053\pm0.010$ & $ $ & $ $&  $0.554$
    \\
  \hline
\end{tabular}
\label{table1}
\end{table*}

Since the witness of quantum states and layered quantum communication protocols only needs two-dimensional subspace projection measurements, we only perform the measurement in two-dimensional subspaces. The measurement device consists of HWPs, QWPs, BDs and PBSs. These setup can also be used to construct measurement setups of any dimension \cite{Guo}.

We have witnessed the fidelity $F_{exp}$ of the ideal quantum state $|\Psi_{442}\rangle$ with the state $\rho_{exp}$. One can conclude that the multipartite entangled state is genuinely (4, 4, 2) entangled from the obtained fidelity $F_{exp}$. This method relies on proving that the measured (4, 4, 2) state cannot be decomposed into entangled states of a smaller dimensionality structure (4, 3, 2). We found the best achievable overlap of a $|\Psi_{432}\rangle$  state with an ideal $|\Psi_{442}\rangle$ state to be $F_{max}$ =0.750 \cite{Supple}. If $F_{exp}>F_{max}$, our state is certified to be entangled in $4\times4\times2$ dimensions. To calculate $F_{exp}$, it is sufficient to measure the 32 diagonal and 6 unique real parts of off-diagonal elements of $\rho_{exp}$ as shown in Fig. \ref{fig3}. Our four-photon counting rate is 0.66/s, and the integration time of each measurement setting is 1800~s. From the experimental data, $F_{exp}$ is calculated to be $0.854\pm0.007$, which is above the bound of $F_{max} =0.750$ by 14 standard deviations. This certifies that the three-photon state is indeed entangled in $4\times4\times2$ dimensions.

In order to prove the layered property of (4, 4, 2) state, we calculate the fidelity of the two-dimensional subspace GHZ state. There are six maximally entangled states in two-dimensional subspaces. Four of them are maximally entangled states of three photons (A, B, C), and two of them are maximally entangled states of two photons (A, B). We still use the fidelity witness to certify the correlation of those subspaces. As shown in Fig.~\ref{fig3}b-g, we measure all diagonal and unique off-diagonal elements of the density matrix. Through them, we can calculate the fidelity of each state and the maximally entangled state. The fidelity of these entangled states are $F= 0.910\pm0.029, 0.906\pm0.030, 0.914\pm0.030, 0.922\pm0.028, 0.933\pm0.030, 0.941\pm0.031$ ($F>0.5$ is the bound for genuinely multipartite entangled states). The results proved that the (4, 4, 2) entangled state we prepared has good correlations in different subspaces.

Our method can be easily extended to generate more-partite high-dimensional layered entangled states. Compared with the OAM DoF in photons, the path DoF is easier to manipulate. According to the method in Ref. \cite{Krenn2}, arbitrary multipartite high-dimensional GHZ quantum states can also be realized by our experimental scheme.

This new type of multipartite high-dimensional entangled quantum state can complete the simplest layered quantum communication network. Due to the dimensionality of four, one can share secret keys among all the three parties, and share secret keys among A and B simultaneously independent of the measurement results of C \cite{Pivoluska}. We take the simplest layered quantum communication network as an example.

\begin{figure}[tbph]
\begin{center}
\includegraphics [width=1\columnwidth]{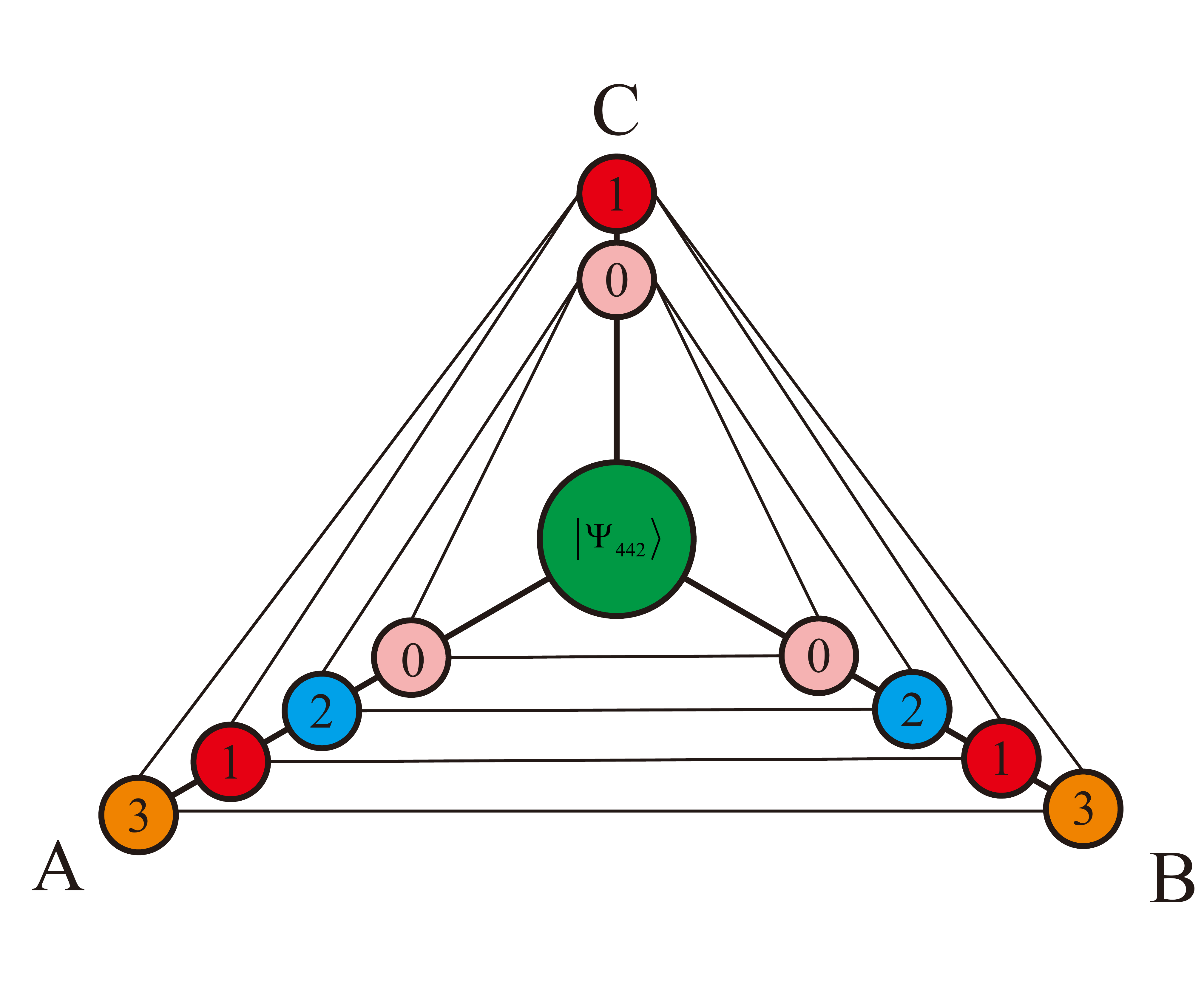}
\end{center}
\caption{A layered quantum communication network. A, B and C share a quantum state $|\Psi_{442}\rangle$, three-party quantum communication or AB two-party quantum communication can be accomplished by using different two-dimensional subspaces.}
\label{fig4}
\end{figure}

As shown in Fig. \ref{fig4}, consider the quantum states (1) we prepared. Each of the four possible outcome combinations $\{000, 111, 220, 331\}$ is distributed uniformly to A, B and C. Moreover, the outcomes of A, B $\{00, 11, 22, 33\}$ are perfectly correlated and partially independent of the outcomes of C. Between A, B and C, the following measurements can be used to complete the tripartite quantum communication $k_{ABC}$.

\begin{equation}
k_{A B C}=\left\{\begin{array}{ll}{0}&{\text{ for outcomes }0\text { and } 2} \\ {1} & {\text { otherwise. }}\end{array}\right.
\end{equation}

At the same time, A and B can communicate with each other in the following way $k_{AB}$.

\begin{equation}
k_{A B}=\left\{\begin{array}{ll}{0} & {\text { for outcomes } 0 \text { and } 1} \\ {1} & {\text { otherwise. }}\end{array}\right.
\end{equation}

$k_{ABC}$ is completely correlated with C's measurement results; therefore, it constitutes a random string shared by the three users. On the other hand, string $k_{AB}$ is completely independent of C's data conditioned on either of C's two measurement outcomes, the value of $k_{AB}$ is 0 or 1, each with probability 1/2. So the simplest layered quantum communication can be achieved by the above method. Of course, we can use the high-dimensional GHZ state or (3, 3, 2) state to complete the same quantum communication protocol, but it is proved that the communication efficiency is lower \cite{Pivoluska}.


In order to assess usefulness of the produced (4, 4, 2) state for quantum key distribution, we calculate asymptotic key rates $R$ for each layer. We use a method developed in \cite{Epping}, which considers security against adversaries using coherent attacks.
More precisely we use their equation (23), which uses parameters $QBER_Z, QBER_X, QBER_Z(AB)$ and $QBER_Z(AC)$.
These are in order: quantum bit error rate between all three parties in $Z$ and $X$ basis, and quantum bit error rate in $Z$ basis between pairs of users $A,B$ and $A,C$.
All of these parameters can be obtained directly from experimental data.
We present values of these experimental parameters for all four layers in table \ref{table1}. Plugging the highest values from these intervals into equation (23) of \cite{Epping} yields a lower bound for the asymptotic key rate per round, which also listed in the table. This is the first demonstration of quantum key distribution utilizing high-dimensional multipartite states.

In conclusion, we have created a (4, 4, 2) quantum state using photons' path and polarization DoFs. This state exhibits different correlations from all the previously reported state \cite{Malik,Erhard,Hiesmayr} because we have promoted the dimension from 3 to 4. The post-selection scheme we employed to increase the dimension allows a heralded generation of the (4, 4, 2) state, with an overhead of 1/2 compared to the canonical (2, 2, 2) generation. We have also experimentally demonstrated, as a proof-of-principle, that this quantum state can complete an efficient layered quantum communication network. Compared with OAM DoF or time bin DoF, the path DoF has higher fidelity and more controllability, so many novel physical phenomena \cite{Jianwei,Hu2016,Hu2} and quantum information tasks \cite{Hu1,Hu2019,Luo} are first realized in the path DoF. Our experimental method can be effectively extended to produce more kinds of multipartite high-dimensional entanglement \cite{Krenn2,Gu}, and to construct more complex high-dimensional quantum networks.

The biggest remaining challenge is the efficiency of promoting bipartite sources of down-conversion into multipartite states. Correlated single-photon sources would provide an obvious route towards more efficient production and would also be compatible with our post-selection scheme.

\begin{acknowledgments}
This work was supported by the National Key Research and Development Program of China (No.\ 2017YFA0304100, No. 2016YFA0301300 and No. 2016YFA0301700), NSFC (Nos. 11774335, 11734015, 11874345, 11821404, 11904357, 61490711), the Key Research Program of Frontier Sciences, CAS (No.\ QYZDY-SSW-SLH003), Science Foundation of the CAS (ZDRW-XH-2019-1), the Fundamental Research Funds for the Central Universities, and Anhui Initiative in Quantum Information Technologies (Nos.\ AHY020100, AHY060300). MH acknowledges funding from the Austrian Science Fund (FWF) through the START project Y879-N27.
MH and MP acknowledge the joint Czech-Austrian project MultiQUEST (I 3053-N27 and GF17-33780L).
\end{acknowledgments}

\clearpage
\onecolumngrid
\begin{center}
{\Large Supplementary Information}\\
\end{center}
\twocolumngrid
\setcounter{table}{0}
\renewcommand{\thetable}{S\arabic{table}}
\setcounter{figure}{0}
\renewcommand{\thefigure}{S\arabic{figure}}
\setcounter{equation}{0}
\renewcommand{\theequation}{S\arabic{equation}}

\section{Dimension increasing device}

\begin{figure}[tbph]
\begin{center}
\includegraphics [width=1\columnwidth]{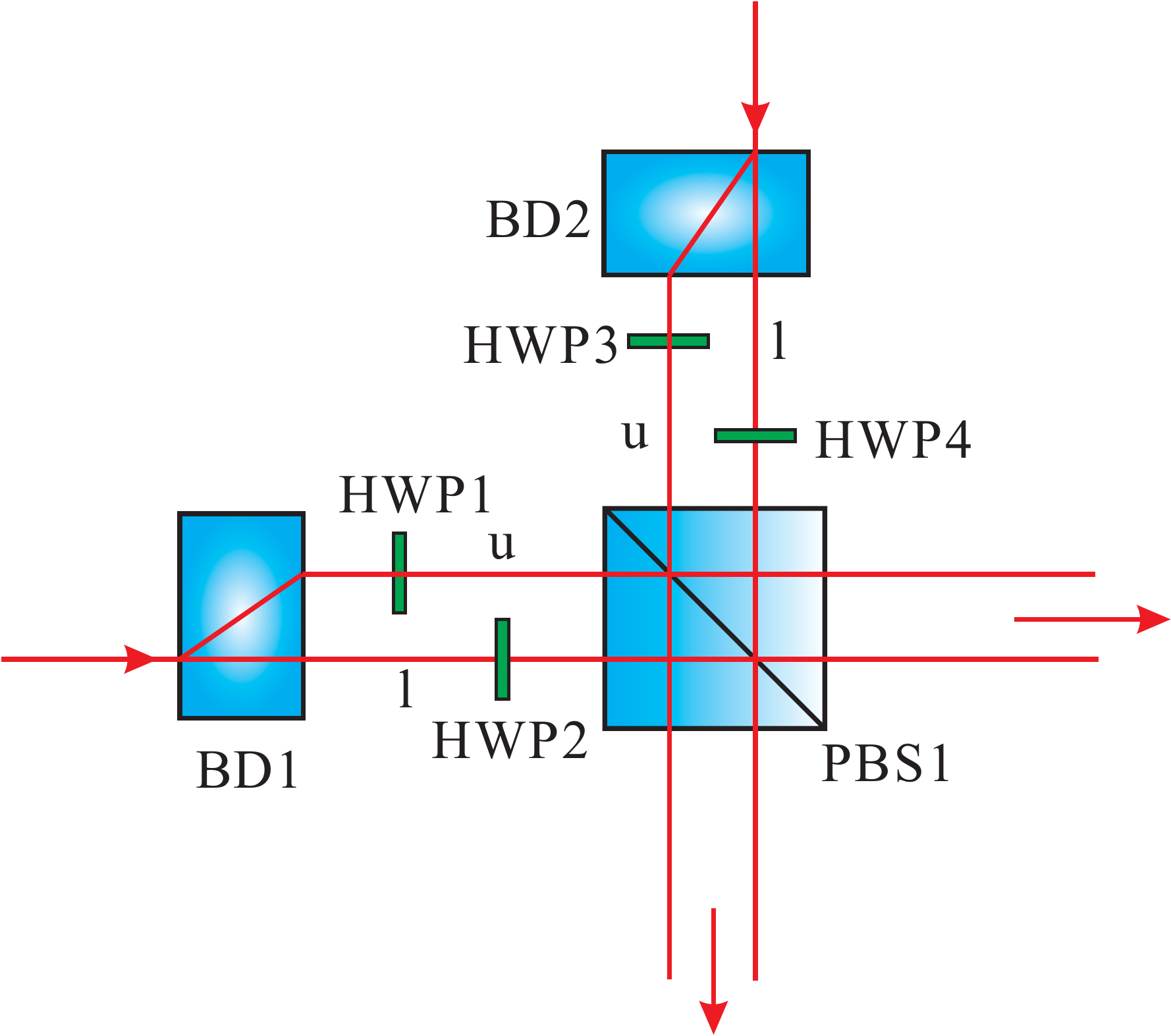}
\end{center}
\caption{Dimension increasing setup.}
\label{figs1}
\end{figure}

In Fig. \ref{figs1}, the main function of this part is to upgrade the two-dimensional entanglement to four-dimensional entanglement through the post-selection method. We consider a two-dimensional entanglement $(|HH\rangle+|VV\rangle)/\sqrt{2}$, which is input to both arms of the device. After the changes of BD1 and BD2, the quantum state becomes a hybrid coding of polarization and path:
\begin{equation}
\frac{1}{\sqrt{2}}(|H_{u}H_{u}\rangle+|V_{l}V_{l}\rangle).
\end{equation}

The angle of HWP1-4 is set at $22.5^{\circ}$ and the quantum state after these HWPs is changed to:
\begin{equation}
\frac{1}{\sqrt{2}}(\frac{1}{2}|H_{u}+V_{u}\rangle|H_{u}+V_{u}\rangle+\frac{1}{2}|H_{l}-V_{l}\rangle|H_{l}-V_{l}\rangle).
\end{equation}

After passing through PBS1, we only keep the item where there is only one photon on both sides
\begin{equation}
\frac{1}{2\sqrt{2}}(|H_{u}H_{u}\rangle+|V_{u}V_{u}\rangle+|H_{l}H_{l}\rangle+|V_{l}V_{l}\rangle).
\end{equation}

If we define $|H_{u}\rangle\longrightarrow|0\rangle$, $|V_{u}\rangle\longrightarrow|2\rangle$, and lower path $|H_{l}\rangle\longrightarrow|1\rangle$ and $|V_{l}\rangle\longrightarrow|3\rangle$. Finally we get the quantum state:
\begin{equation}
\frac{1}{2}(|00\rangle+|11\rangle+|22\rangle+|33\rangle).
\end{equation}

Then the two-dimensional entanglement is chosen to be a four-dimensional entanglement with a probability of $50\%$. Our method can be applied to double the dimension of entanglement in any dimension.

\section{Entanglement witness}

If we want to reconstruct the density matrix of our prepared quantum state, we need to perform 1024 projection measurements, which is a great challenge for multipartite quantum states. So we use the method in \cite{Malik} to witness the entanglement type of our quantum state. To prove that this state is a (4, 4, 2) entangled state, we only need to measure 32 diagonal and 6 unique real parts of off-diagonal density matrix elements. To prove that the state is indeed a (4, 4, 2)-type entangled state, we have to prove that it cannot be decomposed into states of a smaller dimensionality structure. We thus have to show that it lies outside the (4, 3, 2) set of states, that is the convex hull of all states that can be decomposed into (4, 3, 2) and (3, 4, 2) states. Through witness, we can get the fidelity $F_{exp}=\operatorname{Tr}\left(\rho_{exp}|\Psi\right\rangle\langle\Psi|)$ between the quantum state and the ideal state $|\Psi\rangle=\frac{1}{2}(|000\rangle+|111\rangle+|220\rangle+|331\rangle)$. We thus need to compare the experimental fidelity with the best achievable fidelity of a (4, 3, 2)-state, that is
\begin{equation}
F_{max}:=\max _{\sigma \in(432)} \operatorname{Tr}(\sigma | \Psi\rangle\langle\Psi |).
\end{equation}

If $F_{exp}>F_{max}$, we can certify that the state has an entangled dimensionality structure of (4, 4, 2). To calculate $F_{max}$, it is useful to observe that it is convex in the set of states, that is, the maximum will always be reachable by a pure state. Furthermore, as the set (4, 3, 2) is the convex hull of (4, 3, 2) and (3, 4, 2), we can write the following:
\begin{equation}
\begin{aligned} F_{max} &=\max _{ | \Phi \rangle \in(432)}|\langle\Psi | \Phi\rangle|^{2} \\
&=\max \left[\max _{ | \Phi \rangle \in 432}|\langle\Psi | \Phi\rangle|^{2}, \max _{ | \Phi \rangle \in 342}|\langle\Psi | \Phi\rangle|^{2}\right].
\end{aligned}
\end{equation}

Now for a fixed rank vector $xyz$, these fidelities can be bounded by noticing that
\begin{equation}
\begin{aligned}
&\max _{ | \Phi \rangle \in x y z}|\langle\Psi | \Phi\rangle|^{2} \leq \min \left[\max _{\operatorname{rank}\left(Tr_{23} | \Phi\right\rangle\langle\Phi |)=x}|\langle\Psi | \Phi\rangle|^{2} , \right. \\
&\max _{\operatorname{rank}\left(Tr_{13} | \Phi\right\rangle\langle\Phi |)=y}|\langle\Psi | \Phi\rangle|^{2}, \quad \max _{\operatorname{rank}\left(Tr_{12} | \Phi\right\rangle\langle\Phi |)=z}|\langle\Psi | \Phi\rangle|^{2} ].
\end{aligned}
\end{equation}

It states that if a state has a Schmidt decomposition across a cut $A | \overline{A}$ given by $|\Psi\rangle=\sum_{i_{0}}^{r-1} \lambda_{i}\left|v_{A}^{i}\right\rangle \otimes\left|v_{\overline{A}}^{i}\right\rangle$, we can compute the maximal overlap with a state of bounded rank across this partition as
\begin{equation}
\max _{\operatorname{rank}\left(\operatorname{Tr}_{\overline{A}} | \Phi\right\rangle\langle\Phi |)=x}|\langle\Psi | \Phi\rangle|^{2}=\sum_{i=0}^{x-1} \lambda_{i}^{2},
\end{equation}
where we assumed ordered Schmidt coefficients, that is, $\lambda_{i} \geq \lambda_{i+1}$. Now, all we need are the coefficients for the Schmidt decomposition for our target state for all three partitions. For $3|12$ they are $\{1/2,1/2 \}$ and for $2|13$ and $1|23$ we get $\{1/4,1/4,1/4,1/4\}$. Inserting these numbers we find that the maximal overlap of the target state (4, 4, 2) with a (4, 3, 2) state is given by
\begin{equation}
F_{\max }=\frac{3}{4}.
\end{equation}

The experimental fidelity $F_{exp} :=\operatorname{Tr}\left(\rho_{\exp}|\Psi_{442}\right\rangle\langle\Psi_{442}|)$ determines which measurements are required. The projector $| \Psi_{442} \rangle\langle\Psi_{442} |$ projects only onto the nonzero diagonal and off-diagonal elements contained in the density matrix $\rho_{\mathrm{exp}}$.

\begin{equation}
\begin{split}
F_{exp}=&\operatorname{Tr}(\rho_{exp}|\Psi_{442}\rangle\langle\Psi_{442}|)\\
       =&\frac{1}{4}(\langle000|\rho_{exp}|000\rangle+\langle111|\rho_{exp}|111\rangle\\
         &+\langle220|\rho_{exp}|220\rangle+\langle331|\rho_{exp}|331\rangle \\
         &+\langle000|\rho_{exp}|111\rangle+\langle111|\rho_{exp}|000\rangle \\
         &+\langle000|\rho_{exp}|220\rangle+\langle220|\rho_{exp}|331\rangle \\
         &+\langle000|\rho_{exp}|331\rangle+\langle331|\rho_{exp}|220\rangle \\
         &+\langle111|\rho_{exp}|220\rangle+\langle220|\rho_{exp}|111\rangle \\
         &+\langle111|\rho_{exp}|331\rangle+\langle331|\rho_{exp}|111\rangle \\
         &+\langle220|\rho_{exp}|331\rangle+\langle331|\rho_{exp}|220\rangle).
\end{split}
\end{equation}

A diagonal element is given by one single projection $\langle i j k|\rho| i j k\rangle=\left(C(i j k) / C_{T}\right)$, with $C_{T} :=\sum_{i=0,1,2,3} \sum_{j=0,1,2,3} \sum_{k=0,1} C(i j k)$, containing all diagonal elements for normalization.
Due to $\langle ijk|\rho|lmn\rangle+\langle lmn|\rho|ijk\rangle=2\mathfrak{R}[\langle ijk|\rho|lmn\rangle]$, out of the 12 off-diagonal elements, only 6 are unique and need to be measured ($\mathfrak{R}[\langle000|\rho|111\rangle]$, $\mathfrak{R}[\langle000|\rho|220\rangle]$, $\mathfrak{R}[\langle000|\rho|331\rangle]$, $\mathfrak{R}[\langle111|\rho|220\rangle]$, $\mathfrak{R}[\langle111|\rho|331\rangle]$, $\mathfrak{R}[\langle220|\rho|331\rangle]$). Note that the last two off-diagonal elements $\mathfrak{R}[\langle111|\rho|331\rangle]$, $\mathfrak{R}[\langle220|\rho|331\rangle]$ are only in a two-partite superposition. Hence, it can be measured in the standard way that two-partite two-dimensional states are usually measured. To measure the other four off-diagonal elements with projective measurements, we decompose them into $\sigma_{x}$ and $\sigma_{y}$ measurements. The real part of each element can be written as
\begin{equation}
\begin{split}
\mathfrak{R}[\langle ijk|\rho| lmn\rangle]=& \left\langle\sigma_{\mathrm{x}}^{\mathrm{i}, \mathrm{l}} \otimes \sigma_{\mathrm{x}}^{\mathrm{j}, \mathrm{m}} \otimes \sigma_{\mathrm{x}}^{\mathrm{k,n}}\right\rangle-\left\langle\sigma_{\mathrm{y}}^{\mathrm{i}, \mathrm{l}} \otimes \sigma_{\mathrm{y}}^{\mathrm{j}, \mathrm{m}} \otimes \sigma_{\mathrm{x}}^{\mathrm{k}, \mathrm{n}}\right\rangle \\
&-\left\langle\sigma_{\mathrm{y}}^{\mathrm{i}, \mathrm{l}} \otimes \sigma_{\mathrm{x}}^{\mathrm{j,m}} \otimes \sigma_{\mathrm{y}}^{\mathrm{k}, \mathrm{n}}\right\rangle-\left\langle\sigma_{\mathrm{x}}^{\mathrm{i}, \mathrm{l}} \otimes \sigma_{\mathrm{y}}^{\mathrm{j}, \mathrm{m}} \otimes \sigma_{\mathrm{y}}^{\mathrm{k}, \mathrm{n}}\right\rangle,
\end{split}
\end{equation}
where $\sigma_{\mathrm{x}}^{a,b}=| a \rangle\langle b|+| b\rangle\langle a|$ and $\sigma_{y}^{a,b}=i | a \rangle\langle b|-i| b\rangle\langle a |$.

From these measurements, the overlap between the generated state $\rho_{exp}$ and the ideal (4, 4, 2) state $|\Psi\rangle$ is calculated to be $F_{exp} = 0.854\pm0.007$. The error is calculated by propagating the Poissonian error in the photon-counting rates by performing a Monte Carlo simulation of the experiment.

To measure the entanglement in two-dimensional subspaces, we also use the fidelity of the states to certify the entanglement. We assume that the maximally entangled state in the two-dimensional subspace is $|\Psi_{subspace}\rangle=(|ijk\rangle+|lmn\rangle)/\sqrt{2}$. The fidelity can be directly calculated as

\begin{equation}
\begin{split}
F_{exp}=&\operatorname{Tr}(\rho_{exp}|\Psi_{subspace}\rangle\langle\Psi_{subspace}|)\\
       =&\frac{1}{2}(\langle ijk|\rho_{exp}|ijk\rangle+\langle lmn|\rho_{exp}|lmn\rangle\\
         &+\langle ijk|\rho_{exp}|lmn\rangle+\langle lmn|\rho_{exp}|ijk\rangle).
\end{split}
\end{equation}

The measurement method of diagonal and off-diagonal elements are the same as that of (4, 4, 2) state. At the same time, we can verify genuine three-photon entanglement of our prepared three-photon GHZ state using the witness: $W_{G}=I/2-|\Psi_{subspace}\rangle\langle\Psi_{subspace}|$. The value of this witness is directly related to fidelity $\operatorname{Tr}(W_{G}\rho_{exp})=F_{exp}-1/2$. We witness the fidelity of entanglement in six subspaces. All entangled states are proved to be genuinely entangled.

\begin{table*}[htbp]
\caption{The density matrix of quantum state $|\Psi_{442}\rangle$ (real part).}
\begin{tabular}{|c|c|c|c|c|c|c|c|c|c|}
  \hline
  $|000\rangle\langle000|$ & $|100\rangle\langle100|$ &  $|200\rangle\langle200|$ & $|300\rangle\langle300|$ & $|010\rangle\langle010|$ & $|110\rangle\langle110|$ & $|210\rangle\langle210|$& $|310\rangle\langle310|$   \\
  \hline
    $0.220\pm0.016$ & $0.003\pm0.002$ &  $0.001\pm0.001$ & $0.000\pm0.000$& $0.000\pm0.000$ & $0.005\pm0.002$ & $0.003\pm0.002$& $0.003\pm0.002$    \\
  \hline
   $|020\rangle\langle020|$ & $|120\rangle\langle120|$ &  $|220\rangle\langle220|$ & $|320\rangle\langle320|$ & $|030\rangle\langle030|$ & $|130\rangle\langle130|$ & $|230\rangle\langle230|$& $|330\rangle\langle330|$   \\
  \hline
    $0.006\pm0.002$ & $0.002\pm0.001$ &  $0.229\pm0.0016$ & $0.000\pm0.000$ & $0.002\pm0.001$ & $0.002\pm0.001$ & $0.003\pm0.002$ & $0.003\pm0.002$    \\
  \hline
 $|001\rangle\langle001|$ & $|101\rangle\langle101|$ &  $|201\rangle\langle201|$ & $|301\rangle\langle301|$ & $|011\rangle\langle011|$ & $|111\rangle\langle111|$ & $|211\rangle\langle211|$& $|311\rangle\langle311|$   \\
  \hline
    $0.004\pm0.002$ & $0.004\pm0.002$ &  $0.007\pm0.003$ & $0.002\pm0.001$ & $0.003\pm0.002$ & $0.242\pm0.017$ & $0.000\pm0.000$ & $0.005\pm0.0024$   \\
  \hline
   $|021\rangle\langle021|$ & $|121\rangle\langle121|$ &  $|221\rangle\langle221|$ & $|321\rangle\langle321|$ & $|031\rangle\langle031|$ & $|131\rangle\langle131|$ & $|231\rangle\langle231|$& $|331\rangle\langle331|$   \\
  \hline
    $0.003\pm0.002$ & $0.002\pm0.001$ &  $0.002\pm0.001$ & $0.003\pm0.0024$ & $0.004\pm0.002$ & $0.001\pm0.001$ & $0.001\pm0.001$ & $0.234\pm0.0017$    \\
  \hline
     $|000\rangle\langle111|$ & $|000\rangle\langle331|$ &  $|111\rangle\langle220|$ & $|220\rangle\langle331|$ & $|220\rangle\langle000|$ & $|111\rangle\langle331|$ & & \\
  \hline
    $0.208\pm0.004$ & $0.198\pm0.004$ &  $0.208\pm0.004$ & $0.206\pm0.004$ & $0.204\pm0.005$ & $0.221\pm0.005$ & &    \\
  \hline
\end{tabular}
\label{table2}
\end{table*}

\section{ Numerical values of the measured density matrix elements}

Detailed results of real part of quantum state elements $|\Psi_{442}\rangle$  are listed in Table \ref{table2}.

The errors are calculated by propagating the Poissonian error in the photon-counting rates by performing a Monte Carlo simulation of the experiment.

\end{document}